\documentclass[journal=jacsat,manuscript=article]{achemso}
\setkeys{acs}{articletitle = true}

\usepackage[version=3]{mhchem} % Formula subscripts using \ce{}
\usepackage{amsmath}  
\usepackage{amsfonts} 
\usepackage{subcaption}
\usepackage{graphicx} 
\usepackage{booktabs}
\usepackage{threeparttable}
\usepackage[font=footnotesize]{caption}
\usepackage{verbatim}
\usepackage{xcolor}
\usepackage{soul}
\usepackage{comment}
\usepackage{xr}
\usepackage{pdflscape}
\usepackage{tikz}
\usepackage{geometry}
\usepackage{dsfont}
\usetikzlibrary{matrix,positioning,decorations.pathreplacing} 
\SectionNumbersOn
\mciteErrorOnUnknownfalse
\makeatletter
\newcommand*{\addFileDependency}[1]{% argument=file name and extension
  \typeout{(#1)}
  \@addtofilelist{#1}
  \IfFileExists{#1}{}{\typeout{No file #1.}}
}
\makeatother

\newcommand*{\myexternaldocument}[1]{%
    \externaldocument{#1}%
    \addFileDependency{#1.tex}%
    \addFileDependency{#1.aux}%
}

\newcommand{\kcalmol}{\mathrm{kcal/mol}}
\newcommand{\invha}{\mathrm{Ha}^{-1}}

\myexternaldocument{SupportingInfo}

\title{Evaluating Multiconfigurational Trials for Accurate Phaseless Auxiliary-Field Quantum Monte Carlo on 3d Transition Metal Complexes}

\author{Hung T. Vuong}
\author{Ankit Mahajan}
\affiliation{Department of Chemistry, Columbia University, 3000 Broadway, New York, NY, 10027}
\author{John L. Weber}
\affiliation{Schr{\"o}dinger Inc., 1540 Broadway, 24th floor, New York, New York 10036, United States}
\author{James Shee}
\email{james.shee@rice.edu}
\affiliation{Department of Chemistry, Rice University, Houston, TX 77005, USA}
\author{David R. Reichman}
\email{drr2103@columbia.edu}
\affiliation{Department of Chemistry, Columbia University, 3000 Broadway, New York, NY, 10027}
\author{Richard A. Friesner}
\email{raf8@columbia.edu}
\affiliation{Department of Chemistry, Columbia University, 3000 Broadway, New York, NY, 10027}
\date{\today}

\begin{document}

\maketitle
\begin{abstract}
In this study, we evaluate multi-configurational trial wave function protocols for phaseless auxiliary field quantum Monte Carlo (ph-AFQMC) on transition metal containing systems.
First, we benchmark vertical ionization potentials for 22 \(3d\) transition metal complexes against published high-accuracy ph-AFQMC values in a double zeta basis set. We then compute the vertical ionization potential for a set of six metallocenes using our best-performing protocol, alongside ph-AFQMC using a configuration interaction singles and doubles (CISD) trial state. We also analyze the performance of canonical coupled-cluster theory with singles, doubles and perturbative triples (CCSD(T)), as well as its local approximation using domain-based local pair natural orbitals (DLPNO-CCSD(T1)) using different reference orbitals.
To reach the complete-basis-set (CBS) limit, we examine several extrapolation schemes and report CBS-limit ph-AFQMC and CCSD(T) values alongside experimental results. We find that ph-AFQMC with the best-performing trial in a triple zeta basis, followed by CBS correction from DLPNO-CCSD(T1) with unrestricted B3LYP reference orbitals, yields small deviations from experiment at modest cost. Using a CISD trial state in ph-AFQMC gives the closest agreement with experiment (errors \(<\) 2 kcal/mol), albeit with lower scalability.
\end{abstract}

\section{Introduction}\label{introduction}
% Electronic structure problem, AFQMC
% Overview of electronic correlation problem - System of interest

Transition metals (TM) are ubiquitous in biological systems and underpin key catalytic functions.\cite{Moeller1980, Andreini2008} 
For example, nitrogenase, an enzyme that fixes nitrogen in cyanobacteria and rhizobacteria, contains the FeMoco \cite{Einsle2020, Hoffman2014, Spatzal2016}, and the oxygen-evolving complex (OEC) of photosystem II, which catalyzes water oxidation, contains an oxo-bridged $\ce{Mn4CaO5}$ cluster with a \ce{Mn3CaO4} cubane core.\cite{Paul2017,Shen2015,Vinyard2013}. 

The catalytic properties of TM complexes are often attributed to the ability to adopt multiple oxidation states due to their incomplete $d$ shells\cite{Shee2021a}.
At the same time, the complicated electronic structure of TM complexes presents a major challenge for calculating their properties from first-principles quantum chemistry.
Density functional theory (DFT), while popular due to its affordability and relative accuracy, is not generally reliable when it comes to transition metal containing systems.\cite{Aoto2017}

Coupled-cluster methodology, notably singles, doubles and perturbative triples (CCSD(T)), considered the gold standard in quantum chemistry, is capable of producing values reaching chemical accuracy (within $1~\kcalmol$ from experiment) for main group organic molecules, but has an unfavorable scaling of $\mathcal{O}(N^7)$\cite{Bartlett2007}, where $N$ is the system size.
Much of the effort in the past few years to scale up coupled-cluster calculations with system size has focused on the development of local approximations, most notably the domain-based local pair natural orbital (DLPNO)\cite{Riplinger2013, Guo2018}, local natural orbital (LNO)\cite{Rolik2013}, and pair natural orbital (PNO)\cite{Ma2018} approaches. 
While these advances have enabled coupled-cluster calculations to be performed on metalloprotein-derived clusters containing hundreds of atoms\cite{Szabo2023}, the accuracy of even canonical CCSD(T) remains unclear for multi-reference systems, a category that includes many transition metal complexes\cite{Aoto2017}. 

%Overview of ph-AFQMC, recent applications to TM systems
Recently, projector Monte Carlo methods, particularly the auxiliary-field quantum Monte Carlo (AFQMC), have emerged as promising approaches for the quantum chemical modeling of molecular systems.
AFQMC is, in theory, able to recover the exact ground state energy of a system by propagating in imaginary time a state that has a finite overlap with the true ground state. 
For molecular systems, using Gaussian basis sets, AFQMC has a favorable scaling of $O(N^3)-O(N^4)$ per sample (roughly similar to DFT but with a larger prefactor)\cite{Motta2018}.
However, like other projector Monte Carlo methods, AFQMC suffers from the fermionic sign/phase problem that degrades the signal-to-noise ratio and increases the computational expense exponentially.\cite{Troyer2005}. 

The phase problem can be avoided by applying a phaseless approximation, resulting in the phaseless variant of AFQMC (ph-AFQMC), at the cost of a bias in the ground state energy\cite{Zhang2003,Motta2018}.
This phaseless bias can be systematically reduced by using more accurate wavefunctions as trials, i.e., trials that are closer to the true ground state, and the method is exact, even under the phaseless approximation, if the exact ground state is used as the trial.
Major efforts in the past few years have been invested in developing and using better trial wavefunctions for ph-AFQMC. 
Due to the potentially significant degree of static correlation in transition metal complexes, multiconfigurational (MC) wavefunctions are often needed to produce energy estimates within chemical accuracy with ph-AFQMC for such systems. 

Popular choices for MC trials are active space-based wavefunctions such as those provided by the complete active space configuration interaction (CASCI) and its self-consistent field variant (CASSCF) approaches, both of which perform full CI in a pre-defined active space. 
However, the exponential scaling with active space size limits the use of CAS methods to a maximum of approximately 20 orbitals in the active space.

Active space wavefunctions have been used as trial wavefunctions in ph-AFQMC on computing transition metal complexes, and yielded promising results\cite{Shee2018,Rudshteyn2020,Rudshteyn2022,Neugebauer2023,Malone2023}. Algorithmic developments\cite{Weber2022,Mahajan2020,Mahajan2021,Mahajan2022} that reduce the scaling of ph-AFQMC with respect to the number of configurations have enabled trial expansions containing thousands to millions of determinants\cite{Mahajan2021,Malone2023}.

%% Difficulty with converging HCISCF
Several of us have recently performed a study that employed ph-AFQMC with HCI and HCISCF wavefunctions in active spaces ranging between 20-40 orbitals as trials for ph-AFQMC, and computed the vertical ionization potential (IP) for 28 3d transition metal
complexes of varying degree of static correlation (called the 3dTMV set).\cite{Neugebauer2023}
In this study, ph-AFQMC/HCISCF IPs are used to assess the performance of canonical CCSD(T) with different reference states, namely Hartree-Fock (HF) or PBE0, for the single reference (SR) and ``borderline'' single reference/multireference (SR/MR) molecules in the double-$\zeta$ def2-SVP basis. 
Here, ph-AFQMC/HCISCF was found to agree with CCSD(T) regardless of the reference for the SR molecules, while for SR/MR complexes, ph-AFQMC/HCISCF agreed with CCSD(T) if appropriate reference orbitals were used in the CC calculations.
In the def2-SVP basis set, the number of basis functions for the 3dTMV complexes is always smaller than 350. 
However, scaling up to larger complexes involving multi-metal centers, the size of the active space that can adequately capture relevant chemical information will significantly increase, and thus converging an accurate HCISCF wavefunction with tight parameters could potentially become very expensive and challenging. 
Therefore, it is natural to consider cheaper alternative options to generate a more scalable MC trial wave function that can be used without compromising the accuracy of ph-AFQMC. 

Here, we evaluate three alternative protocols for generating MC trials for ph-AFQMC and compare the resulting ph-AFQMC IPs with the published ph-AFQMC values obtained with HCISCF trials. 
In the first protocol, a CASSCF wavefunction is generated in a small active space no larger than (18e, 18o). 
The second protocol, called HCIx2, generates an HCI wavefunction, with the active space selected from the natural orbitals from a larger active space HCI calculation. 
Finally, the third protocol, dubbed HCI-CASSCF, uses the CASSCF-optimized orbitals to generate an HCI wavefunction in a larger active space. 
The latter two protocols produce trials with active spaces comparable to those used for the HCISCF trials in the published 3dTMV studies.
We find that all three protocols outperform single determinant trials, with the third protocol (HCI-CASSCF) yielding the best agreement with the published ph-AFQMC/HCISCF IPs.

Building on the small basis set results, we test the third protocol in a larger triple-zeta basis set and at the complete basis set (CBS) limit by computing the vertical IPs for a set of six metallocenes\cite{Rudshteyn2022}, for which experimental data is available.
We compare our results to those found from ph-AFQMC using CISD trial states,\cite{Mahajan2024} as well as DLPNO-CCSD(T1), and canonical CCSD(T) calculations. 
We find that both HCI-CASSCF and CISD trial wave functions yield results that agree reasonably well with experimental values, given both the experimental and ph-AFQMC error bars, with ph-AFQMC/CISD achieving the smallest error compared to experimental IPs.
Canonical CCSD(T) using UHF reference orbitals yields large discrepancies with ph-AFQMC/CISD due to significant spin contamination, with differences reaching 6 kcal/mol. 
Using ROHF or UB3LYP references, free of or less prone to spin contamination, significantly reduces the difference between ph-AFQMC and CCSD(T) IPs.
DLPNO-CCSD(T1) exhibits substantial local errors with UHF reference orbitals, which can be reduced to acceptable levels by using UB3LYP orbitals instead. 
For CBS extrapolation, we assess four low-level schemes and find that AFQMC/small-CAS and DLPNO-CCSD(T1)/B3LYP perform best. Taken together, these results suggest that HCI-CASSCF provides a more scalable complementary approach to HCISCF, especially when converging HCISCF is infeasible. 
The use of DLPNO-CCSD(T1)/B3LYP for extrapolation to the CBS limit also appears promising, achieving errors of only a few kcal/mol at acceptable computational cost. 

This paper is organized as follows: In Section~\ref{methods}, we discuss the datasets and the details of the MC trial protocols used for ph-AFQMC. 
Additional computational details of ph-AFQMC, DLPNO-CCSD(T1), and canonical CCSD(T), as well as protocols for complete basis set extrapolation, are also presented. 
In Section~\ref{results}, we first present the results of ph-AFQMC for the 3dTMV dataset using the MC trial protocols.
Next, we compare the performance of ph-AFQMC, canonical CCSD(T), and DLPNO-CCSD(T1) IPs for the metallocenes.
We also discuss different CBS extrapolation protocols and compare the results with experimental values.

\section{Methods}\label{methods}
\subsection{Datasets}
In this study, 22 molecules from the original 3dTMV set\cite{Neugebauer2023} are selected, complexes 01-22. 
These complexes are categorized into single reference (SR, complex 01 to 12) and ``borderline" single reference/multireference (SR/MR, complex 13 - 22) subsets based on the MR diagnostics presented in Ref.~\citenum{Neugebauer2023}. 
For these molecules, vertical IPs are computed using ph-AFQMC/HCISCF with active spaces ranging from 20 to 40 orbitals. The $\epsilon_1$ parameter controlling HCISCF accuracy\cite{Smith2017}, is set to $10^{-4}$ a.u.
The ph-AFQMC/HCISCF IPs are used as a reference to benchmark canonical CCSD(T) using either Hartree Fock or Kohn-Sham (PBE) reference orbitals. 
For the SR molecules, CC and ph-AFQMC agree well regardless of the reference orbitals used for CCSD(T), while for the SR/MR molecules, CCSD(T) shows a slight dependence on the reference state, and a qualitatively poor choice can yield IPs deviating by more than $3~\kcalmol$ from ph-AFQMC. 
Complexes 23 to 28 are excluded because they are classified as multireference and show large discrepancies between CCSD(T) and ph-AFQMC IPs of up to $30~\kcalmol$ in some cases. 
Therefore, we include only complexes 01 to 22 (the SR and SR/MR subsets), for which both CC and ph-AFQMC/HCISCF can serve as reference values for evaluating the MC trial protocols tested here.
Hereafter, 3dTMV refers to the combined SR and SR/MR subset, and the MR complexes are not included.
In addition to the 3dTMV dataset, we also select six metallocenes, \ce{MCp2}, where \ce{M} are \ce{V}, \ce{Cr}, \ce{Mn}, \ce{Fe}, \ce{Co} and \ce{Ni}, from Ref.~\citenum{Rudshteyn2022}, and compute their vertical ionization potentials. %In Ref.~\citenum{Rudshteyn2022}, the adiabatic IPs were also computed, but here we focus on vertical ionization for consistency with the 3dTMV dataset. Additionally, the adiabatic ionization is more challenging since it involves changes in the equilibrium geometry of the ionized state, a subject for a future studies. 

\subsection{AFQMC MC Trial Generation Protocols}
The following MC trial generation protocols are benchmarked against ph-AFQMC/HCISCF and canonical CCSD(T) for the 3dTMV dataset:
\begin{itemize}
    \item Protocol (1) - CASSCF: Perform a CASSCF calculation using a small active space no larger than (18e, 18o). Active spaces are chosen based on the RHF and RB3LYP orbital energies.
    \item Protocol (2) - HCIx2: First, run a ``loose" HCI calculation in a relatively large active space ($50-100$ orbitals) with $\epsilon_1=10^{-3}$, starting from RHF or RB3LYP orbitals, to obtain the natural occupation numbers (NOONs). 
    Next, select a smaller set of orbitals by retaining natural orbitals with NOONs in $[\alpha, 2 - \alpha]$.
    For the 3dTMV set, we find that the threshold $\alpha = 0.005$ is sufficient to yield active space sizes comparable to the HCISCF trials in Ref.~\citenum{Neugebauer2023}.
    Finally, perform a ``tight" HCI calculation in this selected active space with $\epsilon_1 = 10^{-4}$ to obtain the trial wavefunction.   
    \item Protocol (3) - HCI-CASSCF: Perform a single-shot HCI calculation with $\epsilon_1 = 10^{-4}$ on the CASSCF optimized orbitals from protocol (1). 
    The HCI active spaces match those used for HCISCF and are likewise based on RHF or RB3LYP orbital energies. 
    This mirrors the protocol used to converge trials for the MR subset of 3dTMV\cite{Neugebauer2023}.
    \end{itemize}

All three protocols are, in principle, cheaper than the original HCISCF approach, since they either use a looser or equal $\epsilon_1$ threshold for HCI, and in protocol (3), perform SCF in a smaller active space than HCISCF. %The final trial wavefunctions generated by all protocols are all in natural orbitals basis. 
Note that protocol (2) does not optimize orbitals in a self-consistent manner; the initial crude HCI calculation serves as an automated way to select the active space for the tighter HCI calculation.  
Details of the active spaces for all three protocols are listed Sec. S1 of the Supporting Information. We also compute vertical IPs of the 3dTMV set with ph-AFQMC/UHF and UPBE0, both single determinant trials. 
For the metallocenes, we adopt the best performing protocol, which is found to be protocol (3) - HCI-CASSCF, to generate the ph-AFQMC trials. 

\subsection{Computational Details of ph-AFQMC}
Trial wavefunctions for ph-AFQMC are generated using PySCF\cite{Sun2020}, and HCI wavefunctions are obtained with DICE\cite{Sharma2017, Smith2017}. 
For the 3dTMV calculations, we use Ahlrichs' def2-SVP basis set\cite{Weigend2005} and correlate all electrons. 
For the metallocenes, we employ Dunning's cc-pVTZ-DKH basis\cite{Dunning1989,Woon1993,Jong2001,Balabanov2005,Balabanov2006} and freeze the \ce{He} core for \ce{C}, and the \ce{Ne} core for the metal.
This leads to the first $15$ orbitals being frozen for each molecule. Scalar relativistic effects are included via the X2C Hamiltonian. 
We use a modified Cholesky decomposition\cite{Motta2018} of the electron repulsion integral (ERI) tensor, with thresholds of $10^{-4}$ a.u. for 3dTMV and $10^{-5}$ a.u. for the metallocenes. %For the 3dTMV dataset, we use a mixed-precision scheme, in which the propagation is done in double precision and the Cholesky matrices are stored in single precision\cite{Neugebauer2023}; while for metallocenes, both propagation and the stored Cholesky matrices are in double precision. 
The propagation time step is $\Delta\tau = 0.005~\invha$ for all calculations. 

For MC trial wave functions, we employ a localized orbital (LO) scheme: the doubly occupied inactive orbitals are localized using Foster-Boys localization, and the half-rotated ERIs are compressed using singular value decomposition (SVD) with a truncation threshold of $5\times10^{-5}$ a.u.\cite{Weber2022}. Control tests reported in Ref.~\citenum{Neugebauer2023} indicate that errors from the mixed-precision scheme and from the Cholesky/SVD thresholds lie within the ph-AFQMC stochastic error bars. Calculations with single determinant trials do not employ this localization scheme. We apply population control using the comb algorithm\cite{Booth2009} every $0.1~\invha$. We use $1728$ walkers and propagate for approximately $200~\invha$ total time. Walkers are orthonormalized every $0.01~\invha$ for numerical stability\cite{Motta2018}, and energy is measured every $0.1~\invha$.

CISD trial wavefunctions for metallocene calculations are obtained by truncating a CCSD wavefunction based on UHF reference orbitals, as detailed in Ref.~\citenum{Mahajan2024}. We used 200 walkers and stochastic reconfiguration for population control. We measure energies every $0.25~\invha$ over a total propagation time of $250~\invha$. We employ the frozen core approximation in these calculations, performed using the \texttt{ad\_afqmc} code\cite{ad_afqmc}.

\subsection{Computational Details of DLPNO-CCSD(T) and CCSD(T)}
Canonical CCSD(T) and DLPNO-CCSD(T) calculations for the metallocenes are performed with the ORCA quantum chemistry package (versions 5.0.3\cite{Neese2022} and 6.0.0\cite{Neese2025}); all DLPNO-CCSD(T) runs use in version 6.0.0. We employ the cc-pVTZ-DKH basis set with the matching auxiliary basis cc-PVTZ/C\cite{Weigend2002,Bross2013}. Scalar relativistic effects are treated with the DKH2 Hamiltonian in ORCA. We use the frozen-core approximation with the same number of frozen orbitals as in the AFQMC calculations. We use DLPNO-CCSD(T) with iterative triples ($T_1$)\cite{Guo2018} using $TCutPNO = 10^{-7}$ and the \texttt{TightPNO} setting, instead of the $T_0$ approximation with \texttt{NormalPNO} used in Ref.~\citenum{Rudshteyn2022}. We perform CCSD(T) and DLPNO-CCSD(T) with RHF, UHF, and UB3LYP reference orbitals.

\subsection{Complete Basis Set Extrapolation of AFQMC}\label{cbs}
Continuum limits of vertical IPs for the metallocenes, computed with ph-AFQMCQMC/HCI-CASSCF and ph-AFQMC/CISD (``high-level'' theories), are obtained by applying TZ/QZ CBS extrapolation corrections from ``low-level'' methods, following procedures similar to those in Refs.~\citenum{Purwanto2011,Shee2019,Rudshteyn2020,Rudshteyn2022}. 
Here we summarize the protocol. For consistency with prior studies, we extrapolate the energy difference, or the ionization potentials, rather than total energies. We consider four low-level schemes for the extrapolation: U-MP2, DLPNO-CCSD(T)/UB3LYP, ph-AFQMC/UHF, and ph-AFQMC/small-CAS. 
The small-CAS trials used in the low-level AFQMC methods are generated using a procedure similar to the ``AutoCAS” protocol for AFQMC-I in Ref.~\citenum{Wei2024}, except that the NOON thresholds for selecting the secondary active space are tightened to 0.02 and 1.98, and the second HCI step is replaced by a CASSCF calculation. This yields consistent active spaces and CI expansions in TZ and QZ, thereby reducing additional sources of extrapolation error. Details of the small CAS trials used can be found in Table S12 of the Supporting Information. 

In the low-level methods, the IP at the CBS limit is given by the sum of the Hartree-Fock (UHF or RHF) IP and the CBS correlation energy as
\begin{equation}
    E_{\infty}^{\mathrm{low}} = E_{\infty}^{\mathrm{HF}} + E^{\mathrm{low}}_{\mathrm{corr}, \infty}.
\end{equation}
The CBS HF IP is obtained by fitting an exponential form as described in Ref.~\citenum{Neese2011}:
\begin{equation}
    E_x^{\mathrm{HF}} = E_{\infty}^{\mathrm{HF}} + A\exp{-\alpha\sqrt{x}},
\end{equation}
where $x = 3$ (TZ) and $x = 4$ (QZ), $\alpha=5.46$\cite{Neese2011}, and $A$ is the fitting constant. The correlation energy is extrapolated to the CBS limit by employing the cubic formula:
\begin{equation}
    E^{\mathrm{low}}_{\mathrm{corr}, \infty} = \dfrac{x^{\beta}E^{\mathrm{low}}_{\mathrm{corr}, x} - y^{\beta}E^{\mathrm{low}}_{\mathrm{corr}, y}}{x^{\beta} - y^{\beta}},
\end{equation}
where $x = 3$ and $y = 4$ for TZ and QZ, and the exponent $\beta=3$\cite{Neese2011}. Once the CBS IP of the low-level method is computed, we obtain the CBS correction $\Delta E^{\mathrm{correction}} = E_{\infty}^{\mathrm{low}} - E_{TZ}^{\mathrm{low}}$, then add this value to the high-level theories TZ results to obtain the final CBS IPs.
\begin{equation}
    E^{\mathrm{high}}_{\mathrm{CBS}} = E_{TZ}^{\mathrm{high}} + \Delta E^{\mathrm{correction}}.
\end{equation}

\section{Results}\label{results}
\subsection{Vertical IPs of 3dTMV SR and SR/MR subsets}
\begin{figure}[H]
    \centering
    \includegraphics[width=0.6\textwidth]{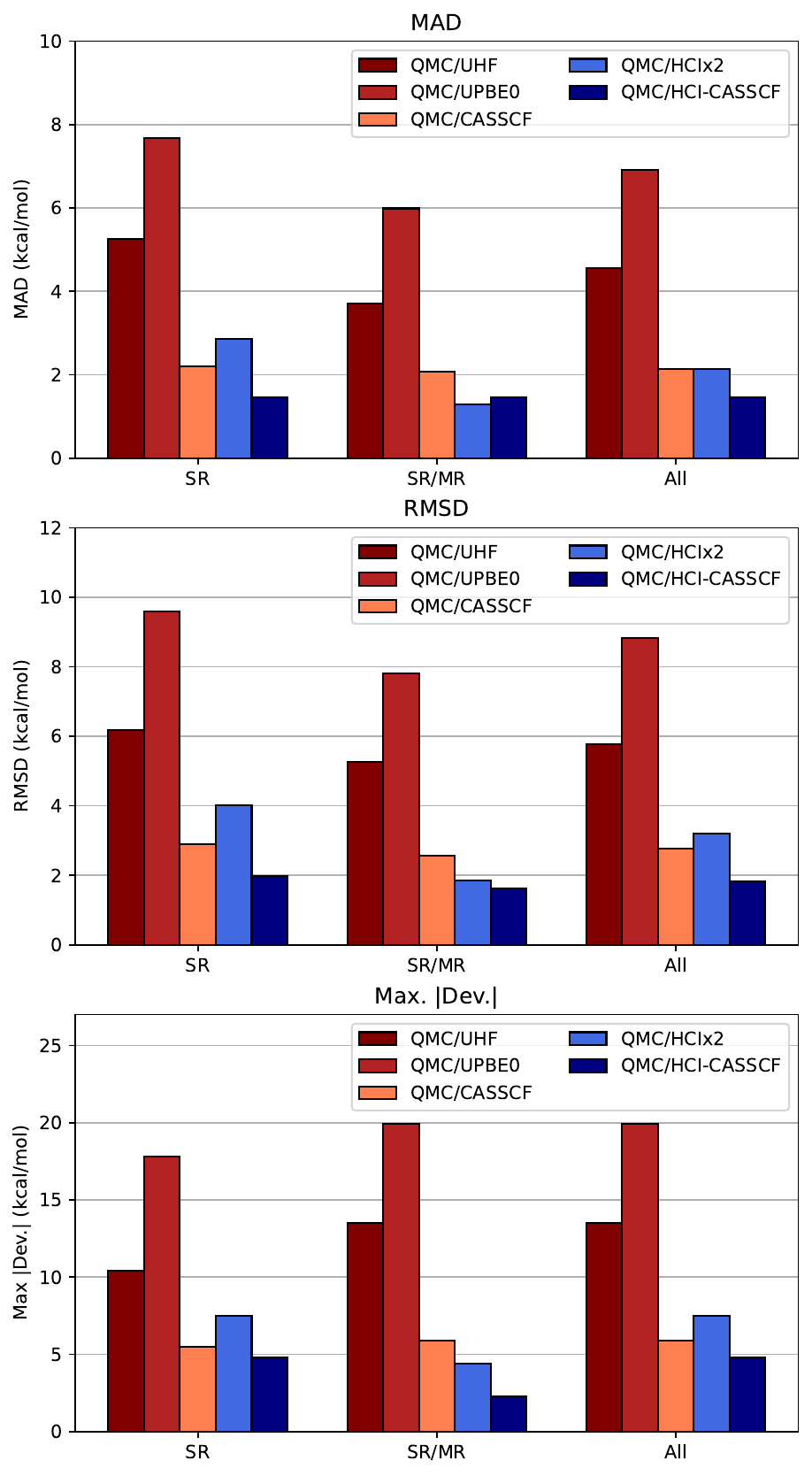}
    \caption{Mean absolute deviation (MAD, top), root-mean-square deviation (RMSD, middle), and maximum absolute deviation (Max $|\mathrm{Dev.}|$, bottom) for ph-AFQMC with different trials for vertical IPs of the SR, SR/MR, and the entire 3dTMV set. ph-AFQMC/HCISCF from Ref.~\citenum{Neugebauer2023} serve as the reference. UHF and UPBE0 are single determinant trials; the remaining three trials are multiconfigurational. ``QMC" denotes ph-AFQMC.}
    \label{fig:01-stats-v-hciscf}
\end{figure}

Fig.~\ref{fig:01-stats-v-hciscf} shows the ph-AFQMC vertical ionization potentials obtained with the MC trial protocols and single determinants trials, along with the LO ph-AFQMC/HCISCF results of Ref.~\citenum{Neugebauer2023}. Results are shown separately for the SR and SR/MR data sets, as well as for the combined data set. Overall, both single determinant trials yield poor agreement with the LO ph-AFQMC/HCISCF references: for the entire set, the MADs are $4.6~\kcalmol$ (UHF) and $6.9~\kcalmol$ (UPBE0), and the RMSDs are $5.8~\kcalmol$ (UHF) and $8.8~\kcalmol$ (UPBE0). The maximum deviations are $13.5~\kcalmol$ (UHF) and $19.9~\kcalmol$ (UPBE0), both exceeding $10~\kcalmol$. Even within the SR subset, performance remains poor (MADs of $6.2~\kcalmol$ for UHF and $7.7~\kcalmol$ for UPBE0), and most SR molecules have errors larger than $4~\kcalmol$ for both trials. Interestingly, ph-AFQMC/UHF performs slightly better for the borderline SR/MR molecules, with MAD of $3.7~\kcalmol$ than for the SR set (MAD $= 5.3~\kcalmol$. This suggests that the MR diagnostics are not good indicators of ph-AFQMC performance. Overall, for TM-containing systems, even many nominally single-reference cases, single-determinant trials are generally insufficient for ph-AFQMC to reach the desired accuracy, and multiconfigurational trials are often required, consistent with prior studies.\cite{Shee2018,Rudshteyn2020,Rudshteyn2022}

Compared with the single determinant trials, all three MC trial approaches substantially improve ph-AFQMC energies. The overall MAD on the 3dTMV set decreases to $2.1~\kcalmol$ for both CASSCF and HCIx2 protocols, with maximum absolute errors of $5.9~\kcalmol$ (CASSCF) and $7.5~\kcalmol$ (HCIx2), and most errors within $3~\kcalmol$. For HCIx2 (protocol 2), the largest errors occur from complexes 8-11 ($\sim6 -7.5~\kcalmol$). Notably, most large-error cases lie in the SR subset (MAD $~2.9~\kcalmol$), rather than SR/MR (MAD $1.3~\kcalmol$). 
Except for complex 18, all SR/MR species show good agreement with the reference ph-AFQMC/HCISCF values using HCIx2. Within the SR subset, the largest-error cases are 8, 9, 10, and 11; aside from 11 (discussed below), these share a common motif: cationic cobalt complexes bearing a cyclopentadienyl (Cp) ligand and phosphine or phosphine-like ligands. Protocol 1 (CASSCF in a small active space) performs comparably for the SR and SR/MR subsets (MADs $=2.4$ and $2.1~\kcalmol$, respectively). Its largest errors occur for complexes 8 ($5.5~\kcalmol$), 9 ($4.7~\kcalmol$), 11 ($5.4~\kcalmol$), 19 ($3.6~\kcalmol$), and 21 ($5.9~\kcalmol$).
Taken together, these results indicate that although both protocols outperform single-determinant trials, the absence of orbital optimization in HCIx2 and the small active-space size in CASSCF can each lead to sizable errors. Thus, a combination of orbital optimization and an adequately large active space is needed to generate reliable trials for ph-AFQMC. 

To combine protocols (1) and (2) while avoiding the cost of HCISCF, we test protocol (3), in which we first obtain a CASSCF wave function in a small active space and then perform a single-shot HCI calculation in a larger active space using the CASSCF-optimized orbitals. This protocol achieves the best agreement with the HCISCF references, with an overall MAD of $1.3~\kcalmol$ for the combined set, the lowest among all trials. The MADs are also consistent across subsets: $1.3~\kcalmol$ for SR and $1.4~\kcalmol$ for SR/MR. Except for complexes 1 ($3.3~\kcalmol$) and 10 ($4.8~\kcalmol$), all IP errors are below $3~\kcalmol$ relative to LO ph-AFQMC/HCISCF. The maximum absolute error is $4.8~\kcalmol$ (complex 10), which is the smallest among the trials tested. We note that for complex 10, the IP obtained from LO ph-AFQMC/HCISCF ($198.5~\kcalmol$) may be unconverged: CCSD(T) with different reference states yields IPs in the range of $200-203~\kcalmol$\cite{Neugebauer2023}, and all three MC trials agree better with CCSD(T) in this case.

We conclude the results for the 3dTMV SR, SR/MR subsets by outlining the pros and cons of each MC trial protocol. Protocol (2) (HCIx2) is the cheapest, since it uses single-shot HCI with relatively ``loose'' $\epsilon_1$ thresholds,\cite{Smith2017} and it substantially improves ph-AFQMC relative to single-determinant trials. However, several complexes still show large deviations from the HCISCF references, and, because no orbital optimization is performed, the quality of the initial RHF or RKS orbitals strongly influences the trial quality. Protocol (1) (small active space CASSCF) also improves upon single determinant trials, but five cases retain considerable errors, suggesting that the results for these cases have not fully converged with respect to the active space. As a cost-effective compromise that combines orbital optimization with a larger active space, protocol (3) (HCI–CASSCF) delivers the best overall agreement with HCISCF, and we test it against experimental IPs for metallocenes in the following section. %A possible drawback of this protocol would be the quality of the first CASSCF wavefunction, for example, if the CASSCF converges to a wrong solution.

\subsection{Vertical Ionization Potential of Metallocenes}
To assess how the MC trials protocols extend to larger basis sets, we compute the vertical IPs for a set of metallocenes. To isolate errors in the treatment of correlation from errors in CBS extrapolation, we first use ph-AFQMC with the best-performing protocol, HCI-CASSCF (protocol 3), as well as with CISD trials in the cc-pVTZ-DKH basis. Additionally, we compute metallocene IPs with canonical CCSD(T) using ROHF, UHF, and UB3LYP reference orbitals. Based on previous benchmarking\cite{Mahajan2024}, we expect AFQMC/CISD to be the most accurate method in this basis and therefore use it as a reference. We then compare various CBS extrapolation schemes by comparing to experimental values. 

\subsubsection{Comparison of ph-AFQMC, CCSD(T) and DLPNO in TZ basis}
\begin{table}[H]
    \centering
    \resizebox{\textwidth}{!}{
    \begin{tabular}{ccccccc}
    \toprule
    Metal & QMC/UHF & QMC/HCI-CASSCF & QMC/CISD & CCSD(T)/UHF & CCSD(T)/UB3LYP & CCSD(T)/ROHF\\
    \midrule
    \ce{V}  & $156.15 \pm 0.62$ &  $155.62 \pm 0.64$ & $154.98 \pm 1.61$ & $153.26$ & $153.40$  & $153.15$\\
    \ce{Cr} & $117.74 \pm 0.51$  & $123.49 \pm 0.71$ & $124.78 \pm 1.69$ & $123.07$ & $124.95$  & $115.13$\\
    \ce{Mn} & $155.84 \pm  0.58$ & $160.51 \pm 0.57$ & $157.28 \pm 1.29$ & $162.27$ & $160.32$ & $159.28$\\
    \ce{Fe} & $157.52 \pm 0.72$ & $157.31 \pm 0.81$ & $156.80 \pm 1.51$ & $155.52$ & $156.61$ & $155.05$\\
    \ce{Co} & $122.62 \pm 0.62$ & $124.52 \pm 0.67$ & $125.59\pm 1.66$ & $122.28$ & $124.63$ & $124.95$\\
    \ce{Ni} & $151.44 \pm 0.55$ & $151.21 \pm 0.66$ & $150.01 \pm 1.69$ & $156.56$ & $150.96$ & $152.30$\\
    \bottomrule
    \end{tabular}}
    \caption{Vertical IPs of metallocenes in cc-pVTZ-DKH basis (in kcal/mol). ``QMC'' denotes ph-AFQMC. ph-AFQMC/UHF values are taken from Ref.~\citenum{Rudshteyn2022}, which correlated all electrons. The other values are obtained using the frozen core approximation.} % All coupled-cluster calculations use DKH2, while QMC values are obtained with x2c.}
    \label{tab:01-mcp2-tz}
\end{table}

The results of the TZ basis calculations are summarized in Table~\ref{tab:01-mcp2-tz}. ph-AFQMC with MC trials (HCI-CASSCF and CISD) produces relatively similar IPs. By contrast, ph-AFQMC using single determinant UHF can differ by as much as $7~\kcalmol$ from ph-AFQMC/CISD (e.g., for \ce{CrCp2}). HCI-CASSCF tracks ph-AFQMC/CISD closely, with a MAD of $1.5~\kcalmol$; five of six IPs are within $3~\kcalmol$ of the ph-AFQMC/CISD values, except \ce{MnCp2} at $3.23~\kcalmol$ (the largest deviation). This consistency across ph-AFQMC trials provides strong validation of our computational approach and increases confidence in the presented results.  

For canonical CCSD(T) with UHF reference orbitals, the MAD and maximum absolute deviation relative to ph-AFQMC/CISD are $3.23~\kcalmol$ and $6.55~\kcalmol$, respectively. The largest discrepancies occur for \ce{MnCp2} ($4.99~\kcalmol$) and \ce{NiCp2} ($6.55~\kcalmol$), whereas the remaining species deviate by $\le 3~\kcalmol$. These two cases, particularly in their charged states, exhibit stronger multireference character than the others,\cite{Rudshteyn2022} 
and \ce{NiCp2} is classified as SR/MR subset in the 3dTMV dataset\cite{Neugebauer2023}. These findings suggest that canonical CCSD(T) may be unreliable for these complexes, with spin contamination likely degrading the accuracy.\cite{Bertels2021,Benedek2022,Neugebauer2023}. Using restricted Hartree-Fock as reference orbitals, i.e., RHF for closed-shell and ROHF for open-shell species, significantly reduces CCSD(T) deviations from ph-AFQMC/CISD values. For five out of six species, CCSD(T)/ROHF differs from ph-AFQMC/CISD by $\le 2.5~\kcalmol$. Most notably, for \ce{MnCp2} and \ce{NiCp2}, CCSD(T)/ROHF agrees much better with AFQMC/CISD than CCSD(T)/UHF, with absolute differences of $2.0~\kcalmol$ and $2.29~\kcalmol$, respectively. The improvement is consistent with spin contamination at the UHF level: $\langle S^2\rangle$ values for the \ce{[MnCp2]+} and \ce{[NiCp2]+} states and for neutral \ce{CoCp2} show large deviations from the ideal values (Table S8), and these cases also show better CCSD(T)–ph-AFQMC agreement when restricted references are used. The only exception is \ce{CrCp2}, where CCSD(T)/ROHF underestimates the IP by $9.65~\kcalmol$, compared to AFQMC/CISD; here, ph-AFQMC/HCI-CASSCF, ph-AFQMC/CISD, and CCSD(T) with unrestricted references are all within $2.0~\kcalmol$ of each other despite the neutral state having a substantial amount of spin contamination at the UHF level (cf. Table S8), consistent with prior def2-SVP results reported in Ref.~\citenum{Neugebauer2023}.

Several studies have suggested that using DFT reference orbitals can enhance the performance of CCSD(T).\cite{harvey2003modelling,beran2003approaching,fang2017prediction} We also find that using Kohn-Sham DFT (UB3LYP) as reference for CCSD(T) improves agreement with ph-AFQMC results. As shown in Table~\ref{tab:01-mcp2-tz}, the maximum absolute deviation with respect to ph-AFQMC/CISD drops to $3~\kcalmol$ (for \ce{MnCp2}) with UB3LYP, compared to $6.55~\kcalmol$ (for \ce{NiCp2}) with UHF. For \ce{NiCp2} specifically, the absolute deviation shrinks to just $0.95~\kcalmol$ when UB3LYP orbitals are used. Overall, CCSD(T)/UB3LYP and ph-AFQMC/HCI–CASSCF perform comparably, with MADs of $1.50$ and $1.34~\kcalmol$, respectively. These findings are consistent with def2-SVP results; for example, the vertical IP of \ce{NiCp2} (complex 14 in the 3dTMV set) decreases by $3.1~\kcalmol$ when using UPBE0 rather than UHF reference orbitals. This improvement can be partly attributed to reduced spin contamination in UB3LYP trials. Collectively, these results indicate that for transition metal complexes with a higher degree of multireference character, such as the charged states of \ce{MnCp2} and \ce{NiCp2}, careful selection of reference orbitals is critical for reliable CCSD(T) predictions. 

\begin{figure}[H]
    \centering
    \includegraphics[width=0.7\linewidth]{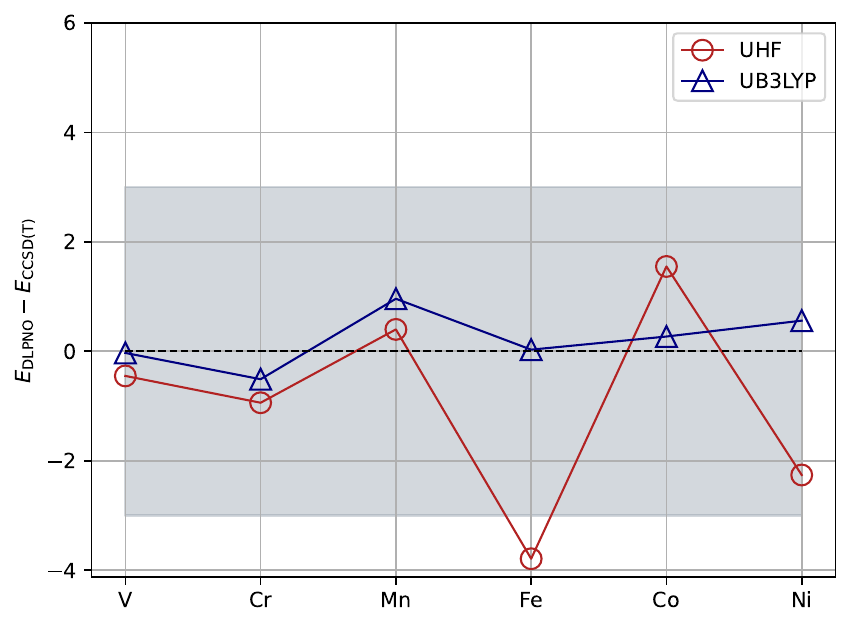}
    \caption{Local error of DLPNO-CCSD(T1) with UHF and UB3LYP reference orbitals for vertical IPs of metallocenes in cc-pVTZ-DKH basis. The shaded gray bar indicates canonical CCSD(T)  $\pm3.0~\kcalmol$.}
    \label{fig2:mcp2_cc_loc_error}
\end{figure}

We conclude our analysis of metallocenes in the TZ basis by examining the DLPNO-CCSD(T1) results. Even with stringent computational thresholds (TCutPNO$ = 10^{-7}$ and \texttt{TightPNO} settings), DLPNO-CCSD(T1) with UHF reference shows sizeable local errors relative to canonical CCSD(T) values (Fig.~\ref{fig2:mcp2_cc_loc_error}). \ce{FeCp2}, and \ce{NiCp2} exhibit particularly large local errors of $-3.79$ and $-2.26~\kcalmol$, respectively. Using UB3LYP reference orbitals for DLPNO-CCSD(T1) significantly reduces these local errors: all metallocenes fall within $1~\kcalmol$, and the local errors for \ce{FeCp2} and \ce{NiCp2} are reduced to $-0.03$, and $0.56~\kcalmol$, respectively. These results suggest that for transition metal complexes, even with the commonly used TightPNO settings, the local error of DLPNO-CCSD(T1) can be substantial and reference-dependent. In some cases, the use of HF orbitals can degrade DLPNO-CCSD(T1) performance, consistent with the conclusions of Ref. \cite{Drabik2024}. Accordingly, we do not extrapolate DLPNO-CCSD(T1)/UHF to the CBS limit and instead extrapolate only DLPNO-CCSD(T1)/UB3LYP for comparison with QMC and canonical CCSD(T).

\subsubsection{Comparison of ph-AFQMC, DLPNO and CCSD(T) at the CBS limit}
To compare with experimental values, extrapolation to the complete basis set limit (CBS) is necessary. The most reliable strategy is to use TZ/QZ extrapolation with high-level calculations (CCSD(T), ph-AFQMC/HCI-CASSCF, and ph-AFQMC/CISD). However, QZ calculations are computationally prohibitive, especially for coupled cluster methods. We also found it challenging to keep the HCI trials consistent between TZ and QZ basis sets, which introduces additional uncertainty in the ph-AFQMC/HCI-CASSCF CBS extrapolations. A more economical CBS extrapolation using lower-level methods is desirable\cite{Drabik2024}. For metallocene vertical IPs, we assess four lower-level theories, viz., UMP2, DLPNO-CCSD(T1)/UB3LYP, ph-AFQMC/UHF, and ph-AFQMC/small CAS (with small active spaces of $8-13$ orbitals), to extrapolate the correlation energy to the CBS limit as described in Sec.~\ref{cbs}. 

\begin{figure}[H]
    \centering
    \includegraphics[width=0.9\textwidth]{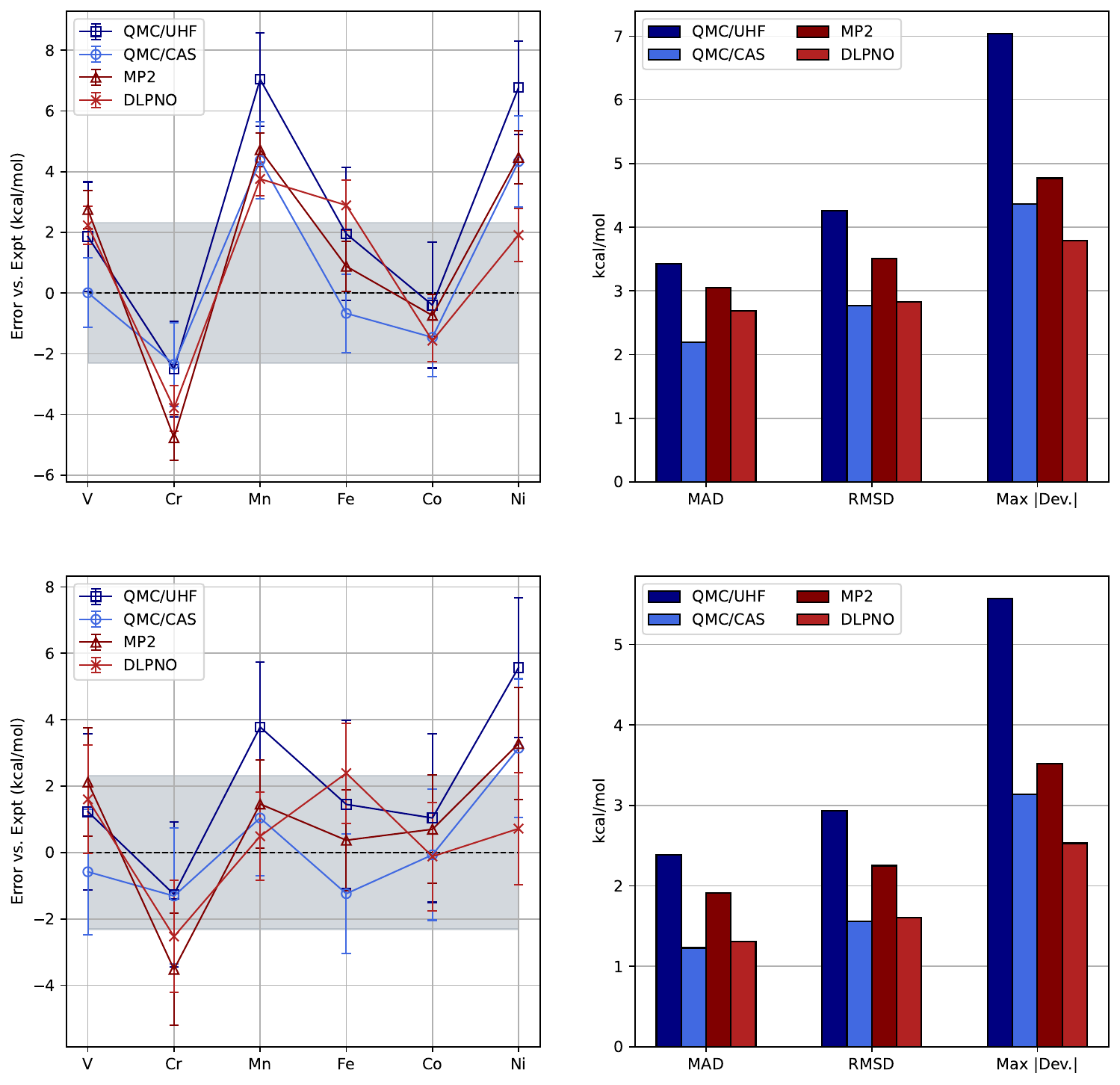}
    \caption{Comparison of CBS corrections, calculated using different low level methods shown in the legends, for metallocene vertical IPs. The top two panels refer to errors (compared to experiment) in CBS-corrected AFQMC/HCI-CASSCF energies, and the bottom ones to errors in CBS-corrected AFQMC/CISD energies. Panels on the left show errors for each metallocene, and those on the right show aggregated errors. The grey bar in the left panels indicates the $\pm 2.31~\kcalmol$ range of experimental error. ``QMC'' is short for ph-AFQMC, and ``DLPNO'' is short for DLPNO-CCSD(T1)/UB3LYP.}
    \label{fig:02-mcp2-cbs}
\end{figure}

The ph-AFQMC results are summarized in Fig.~\ref{fig:02-mcp2-cbs}. ph-AFQMC/CISD with DLPNO-CCSD(T1)/UB3LYP CBS corrections provides the best agreement with experimental values, with all computed values within experimental error bars (2 kcal/mol). UMP2 and ph-AFQMC/small-CAS also yield reasonably accurate corrections on average, but with slightly larger maximum deviations. ph-AFQMC/UHF corrections lead to largest deviations with a maximum error of over 5 kcal/mol. We note that ph-AFQMC based extrapolation corrections contain additional stochastic noise, requiring more sampling to obtain the desired precision. Given the similarity of ph-AFQMC/HCI-CASSCF and ph-AFQMC/CISD energies in the TZ basis, the respective CBS corrected energies display very similar trends. DLPNO-CBS corrected ph-AFQMC/HCI-CASSCF IPs show the smallest deviations from experiment, with an RMSD of around 3 kcal/mol.

\begin{figure}[H]
    \centering
    \includegraphics[width=0.9\linewidth]{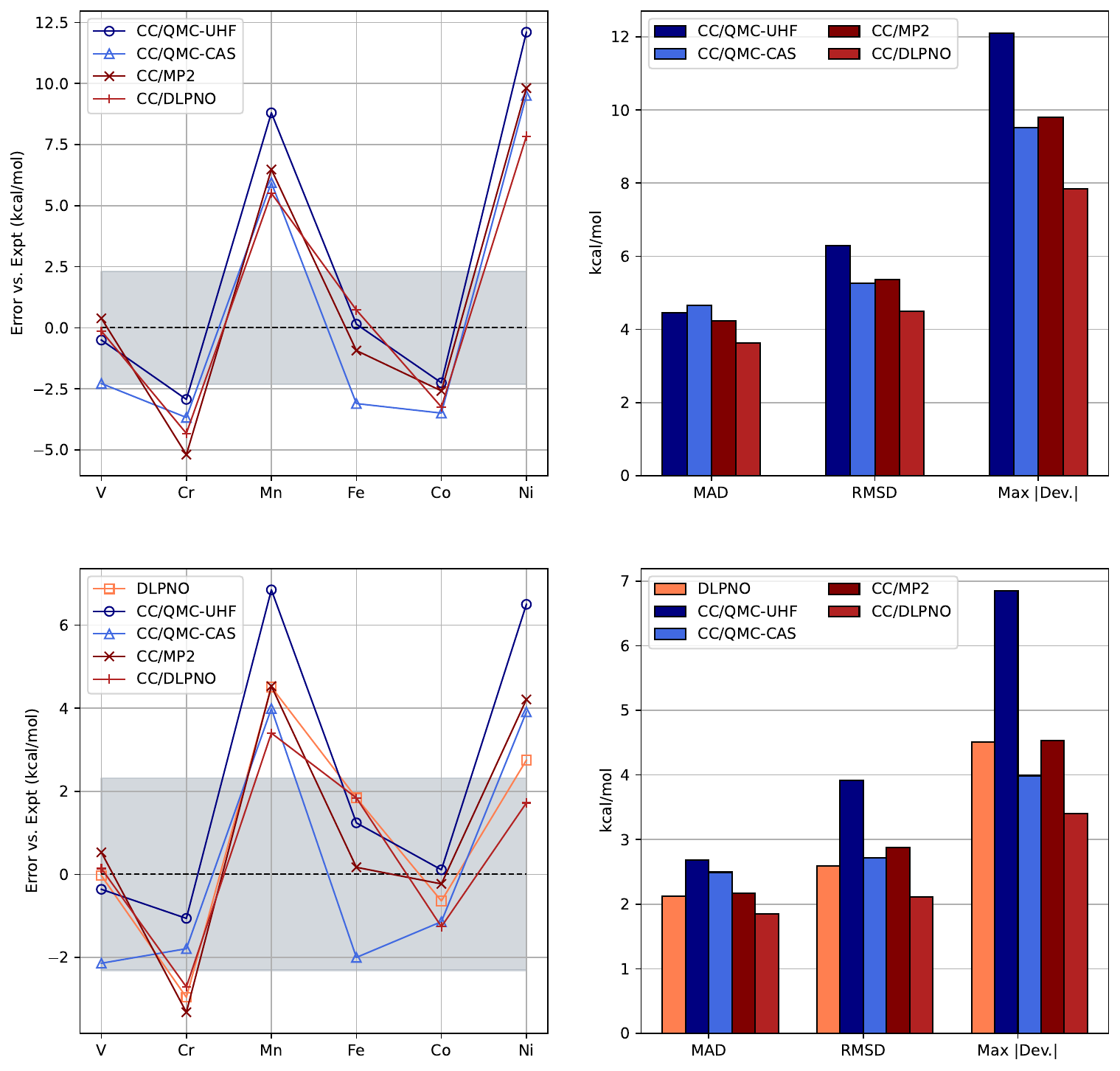}
    \caption{Vertical IPs of metallocenes at the CBS limit from DLPNO-CCSD(T1) and canonical CCSD(T) using UHF reference orbitals (top 2 panels) and UB3LYP reference orbitals (bottom 2 panels). 
    The canonical CCSD(T) values are extrapolated with the indicated low-level methods (“CC/low method” in the legend), whereas DLPNO-CCSD(T1) values use direct TZ/QZ extrapolation. 
    The grey band in the left panels indicates the experimental uncertainty ($\pm 2.31~\kcalmol$). 
    “QMC” denotes ph-AFQMC, “CC” denotes canonical CCSD(T), and “DLPNO” denotes DLPNO-CCSD(T1)/UB3LYP.}
    \label{fig:03-mcp2-cbs-cc}
\end{figure}

CBS-corrected IPs using canonical and DLPNO-CCSD(T1) are compared against experiment in Fig.~\ref{fig:03-mcp2-cbs-cc}. Consistent with the TZ-basis findings, using UHF reference orbitals for CCSD(T) yields poor agreement with experiment regardless of the extrapolation scheme.
The largest errors occur for \ce{MnCp2} (all deviations $>5~\kcalmol$) and \ce{NiCp2} (all deviations $>7.5~\kcalmol$). Using UB3LYP reference orbitals for canonical CCSD(T) improves the results substantially (Fig.~\ref{fig:03-mcp2-cbs-cc}). Among the low-level extrapolation methods, DLPNO-CCSD(T1)/UB3LYP achieves the lowest MAD ($1.8~\kcalmol$), RMSD ($2.1~\kcalmol$), and Max $|\mathrm{Dev.}|$ ($3.5~\kcalmol$) when used to extrapolate canonical CCSD(T)/UB3LYP.
The performance of CCSD(T)/UB3LYP is similar to that of ph-AFQMC/HCI–CASSCF, with the largest deviation from experiment for \ce{MnCp2} ($3.5~\kcalmol$ for CCSD(T)/UB3LYP vs.\ $3.7~\kcalmol$ for ph-AFQMC/HCI–CASSCF). Finally, DLPNO-CCSD(T1)/UB3LYP performs comparably to CCSD(T)/UB3LYP, as expected given the small DLPNO local errors observed in the TZ basis.

A number of conclusions can be drawn from the above results. Firstly, DLPNO-CCSD(T1)/UB3LYP appears to be the best low level extrapolation method, combining high accuracy with a reasonable computational prefactor and scaling with system size. Secondly, the ph-AFQMC/CISD approach followed by DLPNO-CCSD(T1)/UB3LYP CBS extrapolation can be viewed as effectively providing chemical accuracy for the metallocene ionization potentials, given the experimental error bars (in fact, it may well be the case that these computational results are closer to ground truth than the experimental data). Thirdly both the DLPNO-CCSD(T1)/UB3LYP and ph-AFQMC/HCI-CASSCF are within a few kcal/mol of both the benchmark theoretical and experimental results, and hence provide a practical, scalable alternative to ph-AFQMC/CISD for large systems, given the $\mathcal{O}(N^6)$ scaling with system size of the latter methodology. This level of accuracy may well be sufficient to select between chemical reaction mechanisms in many cases. 

A major challenge that remains is how to address large transition metal containing systems with increasingly significant MR character. For cases where the MR impact is unclear, an initial approach is to run both DLPNO-CCSD(T1)/UB3LYP and ph-AFQMC/HCI-CASSCF calculations and compare the results; if they are close, then it is likely that both are correct. If there is a substantial divergence, increasing the size of the ph-AFQMC trial function determinant expansion is an option (although it remains to be demonstrated that such an approach can converge the wavefunction at an affordable computational cost for highly MR cases). 

\section{Conclusions} \label{conclusions}
The accurate computation of properties for transition metal containing systems remains a major challenge in quantum chemistry, and ph-AFQMC is a promising method to achieve high accuracy with favorable scaling. The phaseless bias can be systematically reduced by improving the trial state. For TM complexes, even nominally single-reference cases such as those in the 3dTMV set, single-determinant trials (UHF/UKS) are often insufficient, and multiconfigurational trials are typically required. Selected-CI methods now make it practical to construct MC trials with active spaces beyond the $\sim$(18e,18o) limit of CASSCF. ph-AFQMC/HCISCF has delivered reference-quality ionization potentials for the SR and SR/MR subsets of 3dTMV. However, as system size grows, converging HCISCF can become difficult and expensive, motivating more scalable alternatives for generating MC trials for ph-AFQMC without sacrificing accuracy.

In this study, we evaluated three alternative approaches for generating MC trials, two of which used HCI as the active space solver, and tested them on vertical ionization potentials for the 3dTMV set in the def2-SVP basis. All three protocols are cheaper than generating HCISCF wave functions in comparably sized active spaces. On 3dTMV, CASSCF (protocol 1) and HCIx2 (protocol 2) reproduced ph-AFQMC/HCISCF for most complexes, with six and five cases, respectively, exhibiting deviations $>3~\kcalmol$. Protocol 3 (HCI–CASSCF), which combined orbital optimization with a larger active space, yielded the lowest errors overall, reducing both the mean and maximum deviations relative to ph-AFQMC/HCISCF. Accordingly, we recommended HCI–CASSCF trials as the default choice for larger transition-metal-containing systems in AFQMC calculations.

Extending these approaches to larger basis sets, we computed vertical ionization potentials for the metallocenes in a TZ basis. ph-AFQMC/HCI–CASSCF agreed well with reference ph-AFQMC/CISD IPs, with a maximum deviation of $3~\kcalmol$. Canonical CCSD(T) exhibited nontrivial dependence on the reference orbitals: mitigating spin contamination in reference determinants proved critical for obtaining accurate IPs. DLPNO-CCSD(T1) suffered from substantial local errors with UHF reference orbitals; using Kohn–Sham (UB3LYP) orbitals significantly reduced these errors. For CBS extrapolation, we assessed four low-level schemes and found that DLPNO-CCSD(T1)/UB3LYP yielded reliable corrections at an affordable cost.

Several further improvements could be explored for the protocols tested in this study. Automated procedures such as automated valence active space (AVAS)\cite{Sayfutyarova2017} could be used to select the initial large active space for the first, “loose” HCI calculation, in place of simple orbital-energy cuts. The NOON thresholds could be fine-tuned to converge AFQMC with respect to active-space size; here, for the selected-CI trials, we targeted wave functions in active spaces comparable in size to those used in HCISCF. Along these lines, alternative starting orbitals for HCI or CASSCF, such as UCCSD or $\kappa$-OOMP2\cite{Rettig2022,Shee2021}, could be tested to seed both the initial guess and the NOON-based active-space selection, analogous to protocol (2). Finally, because we focused on vertical ionization potentials (where error cancellation can be significant), future work should assess AFQMC with MC trials for properties that involve geometry changes, such as reaction energies and adiabatic spin and charge gaps.
 
\section{Acknowledgments}

This research used resources of the Oak Ridge Leadership Computing Facility at the Oak Ridge National Laboratory, which is supported by the Office of Science of the U.S. Department of Energy under Contract DE-AC05-00OR22725. 
This work used the Extreme Science and Engineering Discovery Environment (XSEDE), which is supported by National Science Foundation grant number ACI-1548562. 
Calculations used the XSEDE resource Expanse at the SDSC through allocation ID COL151. 
JLW was funded in part by the Columbia Center for Computational Electrochemistry (CCCE). 
JS acknowledges support from the Robert A. Welch Foundation, Award No. C-2212.
A.M. and D.R.R. were partially supported by NSF CHE-2245592. Some calculations were performed on the Delta system at the National Center for Supercomputing Applications through allocation CHE230028 from the Advanced Cyberinfrastructure Coordination Ecosystem: Services and Support (ACCESS) program, which is supported by National Science Foundation grants \#2138259, \#2138286, \#2138307, \#2137603, and \#2138296.
H.V. also thanks Benjamin Rudshteyn for helpful discussions.
This article is reproduced in part with permission from Chapter 3 of Ref.~\citenum{Vuong2025}.

\section{Associated Content}
The supporting information includes: 
\begin{itemize}
\item Data for the active spaces for trials used in three protocols on the 3dTMV test set.
\item Data of ph-AFQMC using the MC trials protocols, including the CI cutoff and number of determinants in the trials, the trial energy, and the total ph-AFQMC energy for each complex.
\item Data of ph-AFQMC using HCI-CASSCF and CISD trials; canonical CCSD(T) and DLPNO-CCSD(T1) energy in cc-pVTZ-DKH basis set for metallocenes.
\item Total energies, small CASSCF trial wavefunctions information, and scaling factors of the low-level methods used in CBS extrapolation of metallocenes.
\item Example ORCA submission script for DLPNO-CCSD(T1) calculations.
\end{itemize}
\newpage
\begin{spacing}{0.85}

\bibliography{./References.bib}
\end{spacing}
\newpage

\end{document}